\documentclass[11pt]{article} 
\pdfoutput=1

\usepackage{amsmath}
\usepackage[norelsize]{algorithm2e}
\usepackage{scalerel}
\usepackage{tikz-cd}
\usepackage{ulem}
\usepackage{float}
\restylefloat{figure}
\usepackage{stackengine}
\usepackage[english]{babel}
\usepackage[light,condensed,math]{anttor}
\usepackage[T1]{fontenc}
\usepackage{kpfonts}
\usepackage{comment}
\usepackage{array}
\usepackage{lmodern}
\usepackage[mathscr]{euscript}
\usepackage{breqn}
\usepackage{todonotes}
\usepackage{slashed}
\usepackage{youngtab}
\usepackage{amsmath}
\usepackage{amssymb}
\usepackage{amsxtra}
\usepackage{color}
\usepackage[margin=1.2in]{geometry}

\newcommand{\iM}{{\mathscr M}}

\usepackage[numbers,sort&compress]{natbib}
\usepackage{hypernat}

\usepackage{hyperref}
\usepackage{xcolor}
\usepackage{graphicx}
\usepackage{lipsum}
\definecolor{titlepagecolor}{cmyk}{1,0,0,.30}
\definecolor{namecolor}{cmyk}{3,1.80,0,.10} 

\newcommand{\bb}{$\blacksquare$}
\newcommand{\wb}{$\square$}

\usepackage[numbers,sort&compress]{natbib}
\usepackage{hypernat}

\usepackage{cleveref}

\makeatletter
\DeclareFontFamily{OMX}{MnSymbolE}{}
\DeclareSymbolFont{MnLargeSymbols}{OMX}{MnSymbolE}{m}{n}
\SetSymbolFont{MnLargeSymbols}{bold}{OMX}{MnSymbolE}{b}{n}
\DeclareFontShape{OMX}{MnSymbolE}{m}{n}{
    <-6>  MnSymbolE5
   <6-7>  MnSymbolE6
   <7-8>  MnSymbolE7
   <8-9>  MnSymbolE8
   <9-10> MnSymbolE9
  <10-12> MnSymbolE10
  <12->   MnSymbolE12
}{}
\DeclareFontShape{OMX}{MnSymbolE}{b}{n}{
    <-6>  MnSymbolE-Bold5
   <6-7>  MnSymbolE-Bold6
   <7-8>  MnSymbolE-Bold7
   <8-9>  MnSymbolE-Bold8
   <9-10> MnSymbolE-Bold9
  <10-12> MnSymbolE-Bold10
  <12->   MnSymbolE-Bold12
}{}

\let\llangle\@undefined
\let\rrangle\@undefined
\DeclareMathDelimiter{\llangle}{\mathopen}%
                     {MnLargeSymbols}{'164}{MnLargeSymbols}{'164}
\DeclareMathDelimiter{\rrangle}{\mathclose}%
                     {MnLargeSymbols}{'171}{MnLargeSymbols}{'171}
\makeatother

 \newcommand{\beq}{\begin{equation}}
                \newcommand{\bea}{\begin{eqnarray}}
                \newcommand{\eea}{\end{eqnarray}}
                 \newcommand{\eeq}{\end{equation}}

\newcommand {\BB}   {\mathbb B}

\newcommand {\BC}   {\mathbb C}

\newcommand {\bH}   {\mathbb H}
\newcommand {\BN}   {\mathbb N}

\newcommand {\BR}   {\mathbb R}

\newcommand {\bR}   {\mathbf{R}}

\newcommand {\qe} {\mathfrak q}

\newcommand {\Hf} {\mathsf{H}}

\newcommand {\ii} {\mathrm{i}}

\newcommand {\be}{\underline{\mathbf{e}}}
\newcommand {\br}{\underline{\mathbf{r}}}
\newcommand {\bn}{\underline{\mathbf{n}}}

\newcommand {\Det} {\tt Det}

\newcommand {\4}{  \mathbf{4}}

\newcommand {\bx}{  \mathbf{x}}
\newcommand {\bX}{ \mathbf{X}}

\newcommand {\BS}   {\mathbb S}
\newcommand {\BT}   {\mathbb T}

\newcommand {\BZ}   {\mathbb Z}

\newcommand {\CalA} {\mathcal A}

\newcommand {\CalC} {\mathcal C}

\newcommand {\CalD} {\mathcal D}

\newcommand {\CalF} {\mathcal F}
\newcommand {\CalG} {\mathcal G}
\newcommand {\CalH} {\mathcal H}
\newcommand {\CalI} {\mathcal I}

\newcommand {\CalL} {\mathcal L}
\newcommand {\CalM} {\mathcal M}
\newcommand {\CalN} {\mathcal N}

\newcommand {\CalP} {\mathcal P}

\newcommand {\CalR} {\mathcal R}
\newcommand {\CalS} {\mathcal S}

\newcommand {\CalU} {\mathcal U}
\newcommand {\CalV} {\mathcal V}
\newcommand {\CalX} {\mathcal X}

\newcommand {\CalZ} {\mathcal Z}

\newcommand{\ve}{\varepsilon}

\renewcommand{\hat}{\widehat}

\newcommand{\Tr}{\mathsf{Tr}\,}

\title{\textbf{Magnificent\ Four}}
\author{Nikita Nekrasov\footnote{Simons Center for Geometry and
    Physics, Stony Brook University, Stony Brook, NY 11794\, , 
    \newline{\tiny on leave of absence from:} 
    Kharkevich IITP RAS, Moscow, ITEP, Moscow .\newline
    e-mail: nnekrasov@scgp.stonybrook.edu}}
\date{}

\begin{document}
\maketitle

\begin{abstract}

We present a statistical mechanical model whose random variables are solid partitions, i.e. Young diagrams built by stacking up four dimensional hypercubes. Equivalently, it can be viewed as the model of random tessellations of ${\bR}^{3}$ by squashed cubes of four fixed orientations. The model computes the refined index of a system of $D0$-branes in the presence of $D8-\overline{D8}$ system, with a $B$-field strong enough to support the bound states. Mathematically, it is the equivariant K-theoretic version of integration over the Hilbert scheme of points on ${\BC}^{4}$
and its higher rank analogues, albeit the definition is {\sl{real-}}, not complex analytic. The model is a mother of all random partition models, including the equivariant Donaldson-Thomas theory and the four dimensional instanton counting. Finally, a version of our model with infinite solid partitions with four fixed plane partition asymptotics is the vertex contribution to the equivariant count of instantons on toric Calabi-Yau fourfolds. 

The conjectured partition function of the model is presented. We have checked it up to six instantons (which is one step beyond the checks of the celebrated P.~MacMahon's failed conjectures of the early $\sl XX$ century). A specialization of the formula is our earlier (2004) conjecture on the equivariant K-theoretic Donaldson-Thomas theory, recently proven by A.~Okounkov \cite{AO15}. 

\end{abstract}

\section{Introduction}

This paper has several facets. From the mathematical point of view we are studying a combinatorial problem. We assign a complex-valued probability to the collections
of hypercubes in dimensions two to four, called the partitions, plane partitions, and solid partitions, respectively, and investigate the corresponding partition functions. From the physical point of view we are studying a bound state problem in the supersymmetric quantum mechanics of the system of point-like particles defined using the $D0$-branes of IIA string theory in a presence of a sophisticated domain wall-type defect, a configuration of the $D8-{\overline{D8}}$-branes. The \emph{Four} in the title refers to the number of complex spatial dimensions of the $D8$-branes. The same four is the number of real euclidean dimensions of the hypercubes forming the solid partitions of the combinatorial problem.  The adjective \emph{Magnificent} reflects this author's conviction that the dimension four is the maximal dimension where the natural albeit complex-valued probability distribution exists.  

The connection between the physical and mathematical problems goes through the definition of a moduli space ${\iM}_{k}$ of solutions to a system of quadratic matrix equations, generalizing, in a certain manner, the ADHM equations \cite{ADHM} encoding the solutions of instanton equations
\beq
F_{A} = - \star F_{A}
\label{eq:asd}
\eeq
in the ordinary four dimensional gauge theory. The classical gauge theory studies the solutions of the partial differential equations 
\beq
D_{A}^{*}F_{A} = 0 \,  ,
\eeq
which describe the critical of the Yang-Mills functional
\beq
S_{\rm YM} = \frac{1}{g^2} \int_{X^{4}} {\Tr} F_{A} \wedge \star F_{A} \, \geq \, \Biggl\vert \, \frac{1}{g^2}  \int_{X^{4}} {\Tr} F_{A} \wedge F_{A} \, \Biggr\vert
\label{eq:est}
\eeq
The Eqs. \eqref{eq:asd} are solved by the absolute minima (cf. \eqref{eq:est}) of the Yang-Mills action in a given topological sector. 
In quantum gauge theory, the path integral
 \beq
 \int_{\left[ {\CalA}/{\CalG} \right]}\, \lbrace {\CalD}A \rbrace \ e^{- S_{\rm YM} + \frac{{\ii}{\vartheta}}{8{\pi}^{2}} \int_{X^{4}} {\Tr} F_{A} \wedge F_{A}}
 \label{eq:ympf}
 \eeq
could be potentially approximated, for small $g^2$, by the saddle point contributions. In reality $g^2$ is not a parameter of the model due to asymptotic freedom. Instead, a more sophisticated version of \eqref{eq:ympf} involving fermions can be sometimes evaluated exactly \cite{N2} using an analogue of Duistermaat-Heckman formula. 

{}In higher dimensions one expects the saddle point approximation  to be even more problematic given the dimensionful nature of the coupling constant, which measures the non-linearity of the corresponding partial differential equations.  However, supersymmetry might come to rescue there as well. With the introduction in \cite{N2} of localization techniques in the exact computations in supersymmetric gauge theories in four and higher dimensions \cite{NJ}, the enumeration of partitions, a more then a century-old subject \cite{MM}, has been revived with a novel set of natural (complex) probability measures. 
The partitions, and their higher dimensional analogues plane and 
solid partitions which we discuss below, enumerate the extrema
of the analogues of the Yang-Mills action defined on noncommutative
spaces. The noncommutative gauge theory arises in a limit \cite{SW3} of string theory, yet may capture some of the non-local features of the latter without all its degrees of freedom. 
Formally one may work with the semi-topological theory  in eight dimensions introduced in \cite{BKS}, and its upgrades to nine dimensions along the lines of \cite{NPhD, BLN}. However the most important question is how one compactifies the moduli space of solutions to the equations 
\beq
\star F_{A} = - T_{4} \wedge F_{A}
\label{eq:gasd}
\eeq
 generalizing \eqref{eq:asd} for the higher dimensional spaces of special holonomy (the closed four-form $T_{4}$ being preserved by that special holonomy). Part of the interest in these solutions is their 
relation to calibrated cycles of \cite{Harvey:1982xk}, in particular special Lagrangian submanifolds in Calabi-Yau fourfolds. 

In the present work we shall be mostly interested in the so-called {\it fat points}, i.e.
the solutions with the action concentrated in codimension eight. Physically these arise in the context of $D0$-branes possibly bound to $D8$-brane(s) wrapping a Calabi-Yau fourfold, with the $B$-field turned on, cf. \cite{W2}. It is interesting to allow for both $D$-branes and anti-branes, as in the $D8$-$\overline{D8}$ configuration of \cite{SS}. However, unlike the \cite{SS} setup we may have supersymmetry restored via tachyon condensation as in
\cite{NPr}. The importance of the $D0$-brane bound state problem \cite{Sav, Yi:1997eg, MNS2} for the non-perturbative string dynamics is well-known \cite{WMth}.

The mathematical challenge of the enumeration of the solutions to \eqref{eq:gasd} is the need to deal with the Pfaffians of the (twisted) Dirac operators arising in the linearized problem, also known as the {\it orientation problem}. We shall discuss this problem below.

\section{Partitions}

\subsection{Ordinary\ partitions}
Partitions enumerate ways of representing a whole as a sum of its parts.  

For a non-negative integer $| {\lambda} |$ its partition $\lambda$ is its representation as a sum of non-negative integers, more specifically a partition $\lambda$ is a non-decreasing sequencee of non-negative integers:
\beq
{\lambda} = ( {\lambda}_{1} \geq {\lambda}_{2} \geq \ldots \geq {\lambda}_{{\ell}({\lambda})} )
\label{eq:ordpart}
\eeq
where 
\beq
| {\lambda} | = {\lambda}_{1} + \ldots + {\lambda}_{{\ell}({\lambda})}
\label{eq:sizpart}
\eeq
is called the size of the partition $\lambda$ and ${\ell}({\lambda})$ is called its length. One represents the partition $\lambda$ with the help of the \emph{Young diagram} which is a collection of neatly packed squares $\square$ of total number $|{\lambda}|$ arranged into a pile of ${\ell}({\lambda})$ rows of lengths ${\lambda}_{1}$, ${\lambda}_{2}$, ... , ${\lambda}_{\ell ({\lambda})}$, as in the picture:

\centerline{\includegraphics[width=10cm]{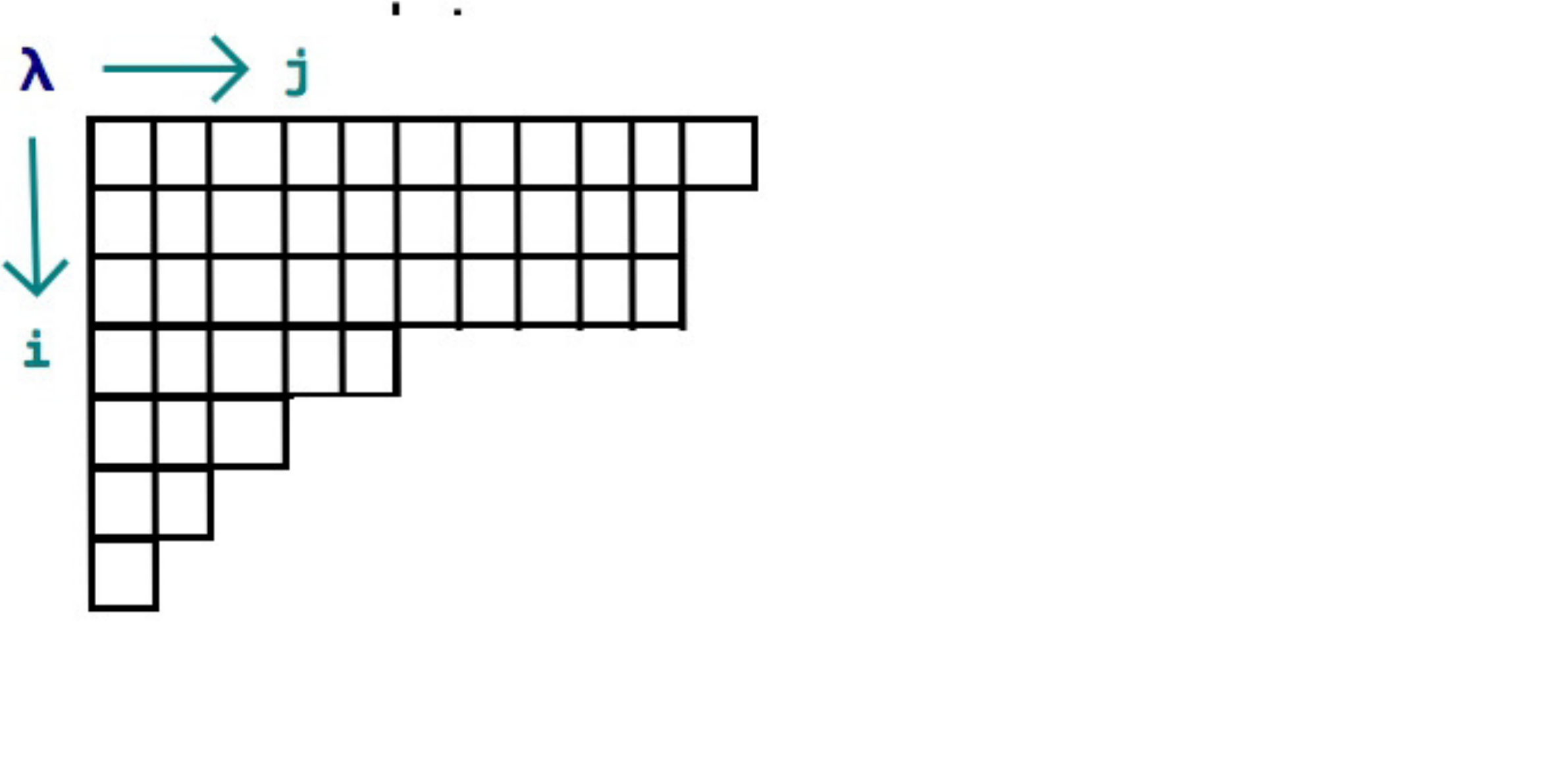}} 

The picture also shows a way of coordinatizing the Young diagrams: we assign to the pair of integers $(i,j)$ obeying: $1 \leq i \leq {\ell}({\lambda})$, $1 \leq j \leq {\lambda}_{i}$ a single square positioned in the $i$'th row counted from the top, and the $j$'th column counted from the left. 

The Young diagrams, i.e. partitions traditionally show up in the representation theory of symmetric groups, for $\lambda$'s are in one-to-one correspondence with the irreducible representations of $S(|{\lambda}|)$. They also enumerate the irreducible representations of the special unitary groups, this time ${\ell}({\lambda})$ plays a role: ${\lambda}$'s are in one-to-one correspondence with the irreducible representations of $SU(N)$ with $N > {\ell}({\lambda})$.

\subsection{Multi-dimensional\ partitions} 

An obvious generalization is to stack cubes, hypercubes etc. Here is a picture\footnote{Thanks to the graphics talents of A.~Okounkov} of the three dimensional Young diagram, also known as the plane partition:

\bigskip
\centerline{\includegraphics[width=8cm]{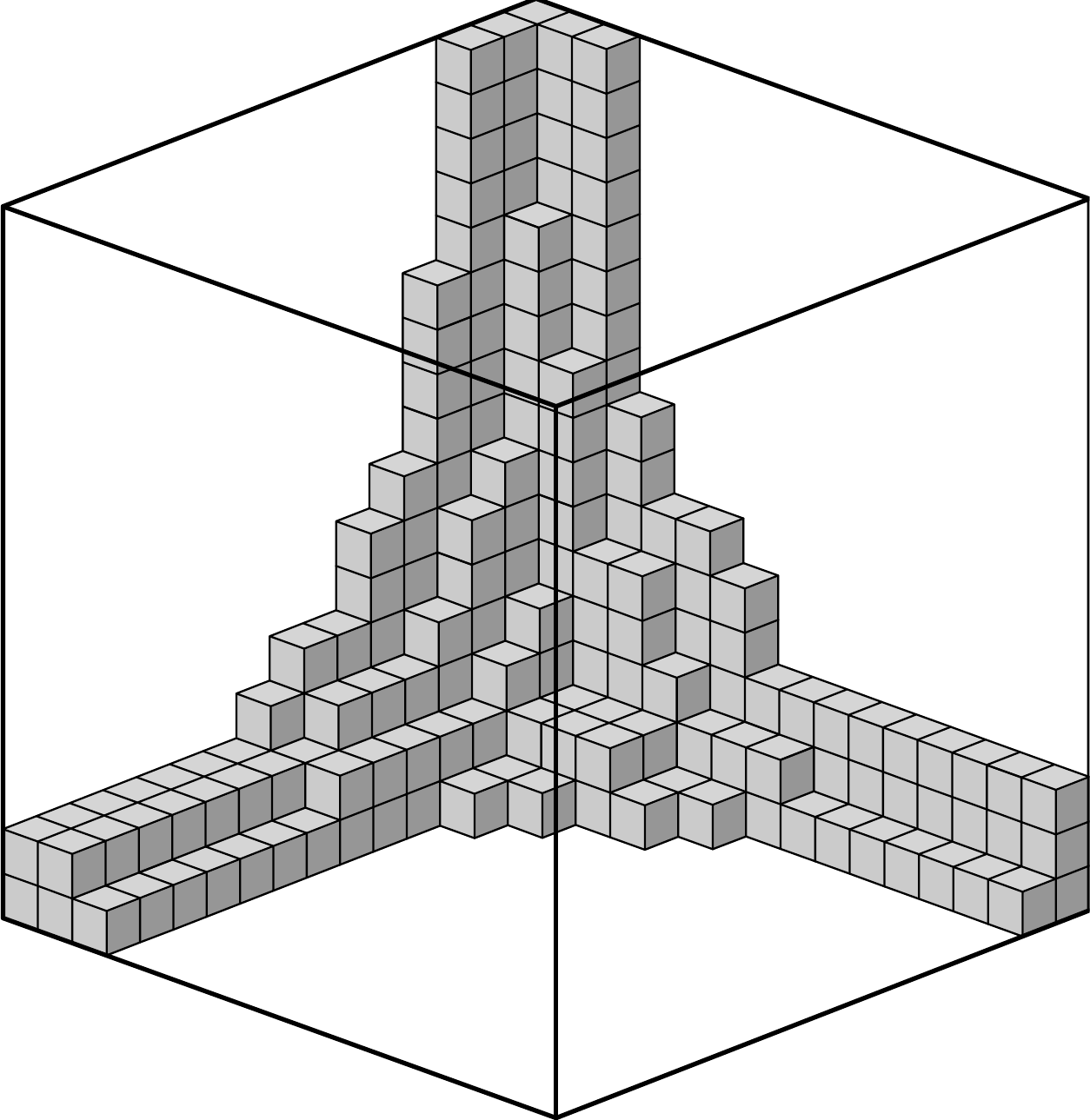}}
\bigskip

More formally, we consider the following generalization of
the structure \eqref{eq:ordpart}. Let $S$ be a
partially ordered set $S$ with valuation $| {\cdot} | : S \to {\BZ}_{\geq 0}$, such that
is $s_{1}, s_{2} \in S$, $s_{1} \geq s_{2}$, then $|s_{1}| \geq |s_{2}|$.  
Define ${\CalP}(S)$ to be the set of finite sequences:
\beq
{\CalP}(S) = \{ \, {\sigma} = \left( s_{1}, s_{2}, \ldots , s_{\ell} \right) \, \vert \, s_{i} \geq s_{j}
\, \ {\rm for\ all} \ i \leq j \, \}
\label{eq:spart}
\eeq 
Then ${\CalP}({\BN}) = {\Lambda}$ is the set of all ordinary partitions, ${\CalP}({\Lambda}) = {\Pi}$ is the set of plane partitions, ${\CalP}({\Pi}) = {\CalS}$ is the set of solid partitions, the
subject of this paper. 

We can view the plane (three-dimensional) partition $\pi$ as a Young diagram ${\lambda}_{\pi}$ with ${\BZ}_{\geq 0}$-valued function ${\pi}_{\square}$ on the set of its squares $\square \in {\lambda}_{\pi}$, with the condition that 
\beq
{\pi}_{i,j} \geq {\pi}_{i+1, j}, \qquad {\pi}_{i,j} \geq {\pi}_{i, j+1}
\eeq
where we write 
 ${\pi}_{i,j} = {\pi}_{\square}$ for the square ${\square} = (i,j) \in {\lambda}_{\pi}$.  The size of the plane partition $\pi$ is defined as
 \beq
 |{\pi} |  = \sum_{i,j} {\pi}_{i,j}
 \eeq
 There
 are three such representations (corresponding to the choice of the labeling of the three coordinate axes). We may call such representations $(2,1)$ types ($2$ for the dimensionality of the Young diagram, $1$ for the fact that we write only a number in each square). Equivalently, we can view $\pi$ as a sequence of non-increasing ordinary partitions. There are also three such representations. We call them $(1,2)$ representations. 
 
 For the solid partitions we can use $(3,1), (2,2)$ or $(1,3)$-pictures: as three
 dimensional Young diagrams with the height functions on the set of cubes (four such representations), 
 as two dimensional Young diagrams with the $\Lambda$-valued function on the set of squares, and as the sequences of non-increasing plane partitions. In the $(3,1)$ representation we fix a plane partition $\pi$ and the height function
 \beq
 {\rho}_{i,j,k} \geq 1, \ (i,j,k) \in {\pi} \, , {\rho}_{i,j,k} \geq {\rho}_{i+1,j,k}, \, {\rho}_{i,j,k} \geq h_{i,j+1,k}, \, {\rho}_{i,j,k} \geq {\rho}_{i,j,k+1} \ . 
 \label{eq:solpart}
 \eeq
 We define ${\rho}_{i,j,k} = 0$ for $(i,j,k) \in {\BN}^{3} \backslash {\pi}$. The size of the solid partition is, naturally
 \beq
 |{\rho}| = \sum_{i,j,k} {\rho}_{i,j,k}
 \eeq
 
 \subsubsection{Enumeration of partitions}
Enumeration of partitions has a long history. The naive question is given the size how many partitions, plane partitions, solid partitions, etc. are there. Equivalently, one is looking for the grand canonical ensemble partition function
\beq
Z_{d} ({\qe}) = \sum_{n=0}^{\infty} p_{d}(n)  \ {\qe}^{n}
\eeq
where $p_{d}(n)$ is the number of $d$-dimensional partitions of size $n$. 

For each $d\geq 1$ the functions
\beq
 {\CalM}_{d}({\qe}) = \sum_{n=0}^{\infty} {\tilde p}_{d}(n)\ {\qe}^{n} = \prod_{n=1}^{\infty} \frac{1}{\left( 1 - {\qe}^{n} \right)^{c_{d}(n)}} \ = \ {\exp} \, \sum_{l=1}^{\infty} \, 
\frac{1}{l} \, f_{d}({\qe}^{l})  
\label{eq:zdq}
\eeq
where 
\beq
c_{d}(n) = \left( \begin{matrix} n+d-3 \\ d-2 \end{matrix} \right) \, , \qquad 
f_{d}({\qe})  = \frac{\qe}{(1-{\qe})^{d-1}}
\eeq
were introduced in \cite{MM}. 
In particular:
\beq
\begin{aligned}
& {\CalM}_{2}({\qe}) = 1 + {\qe} + 2{\qe}^{2} + 3{\qe}^{3} + 5{\qe}^{4} + 7 {\qe}^{5} + 11{\qe}^{6} + \ldots  \\
& {\CalM}_{3}({\qe}) = 1 + {\qe} + 3{\qe}^{2} + 6{\qe}^{3} + 13{\qe}^{4} + 24 {\qe}^{5} + 
48{\qe}^{6} + \ldots  \\
& {\CalM}_{4}({\qe}) = 1 + {\qe} + 4{\qe}^{2} + 10{\qe}^{3} + 26{\qe}^{4} + 59 {\qe}^{5} +
 141{\qe}^{6} + \ldots  \\
\end{aligned}
\label{eq:234}
\eeq
 $Z_{d}({\qe}) = {\CalM}_{d}({\qe})$ was conjectured by P.A.~MacMahon \cite{MM} {\it as part of the  
``... author's preliminary 
researches in combinatory theory which have been carried
out during the last thirty years''}.
\hbox{\vbox{\hbox{The generating functions ${\CalM}_{2}, {\CalM}_{3}$}
\hbox{actually do count partitions and plane partitions,}
\hbox{respectively. For $d=2$, $Z_{2} = {\CalM}_{2}$ is Euler's formula,}
\hbox{which in modern language relates free bosons}
\hbox{and free fermions in $1+1$ dimensions.}
\hbox{The $d=3$ formula has many proofs, one of them}
\hbox{also uses free fermions \cite{OR}.}
\hbox{For some strange reason ${\CalM}_{3}$ coincides with}
\hbox{the partition function of free conformally coupled}
\hbox{scalar on ${\BS}^{1} \times {\BS}^{2}$, with ${\qe} = e^{-R_{1}/R_{2}}$.}
\hbox{For $d=4$, ${\CalM}_{4} - Z_{4} = 0 ($ {\rm mod} ${\qe}^6 )$.}
\hbox{But the conjecture is false, giving ${\tilde p}_{4}(6) = 141$} 
\hbox{instead of the true value $p_{4}(6) = 140$}
\hbox{of solid partitions of $6$:}}
\includegraphics[width=6cm]{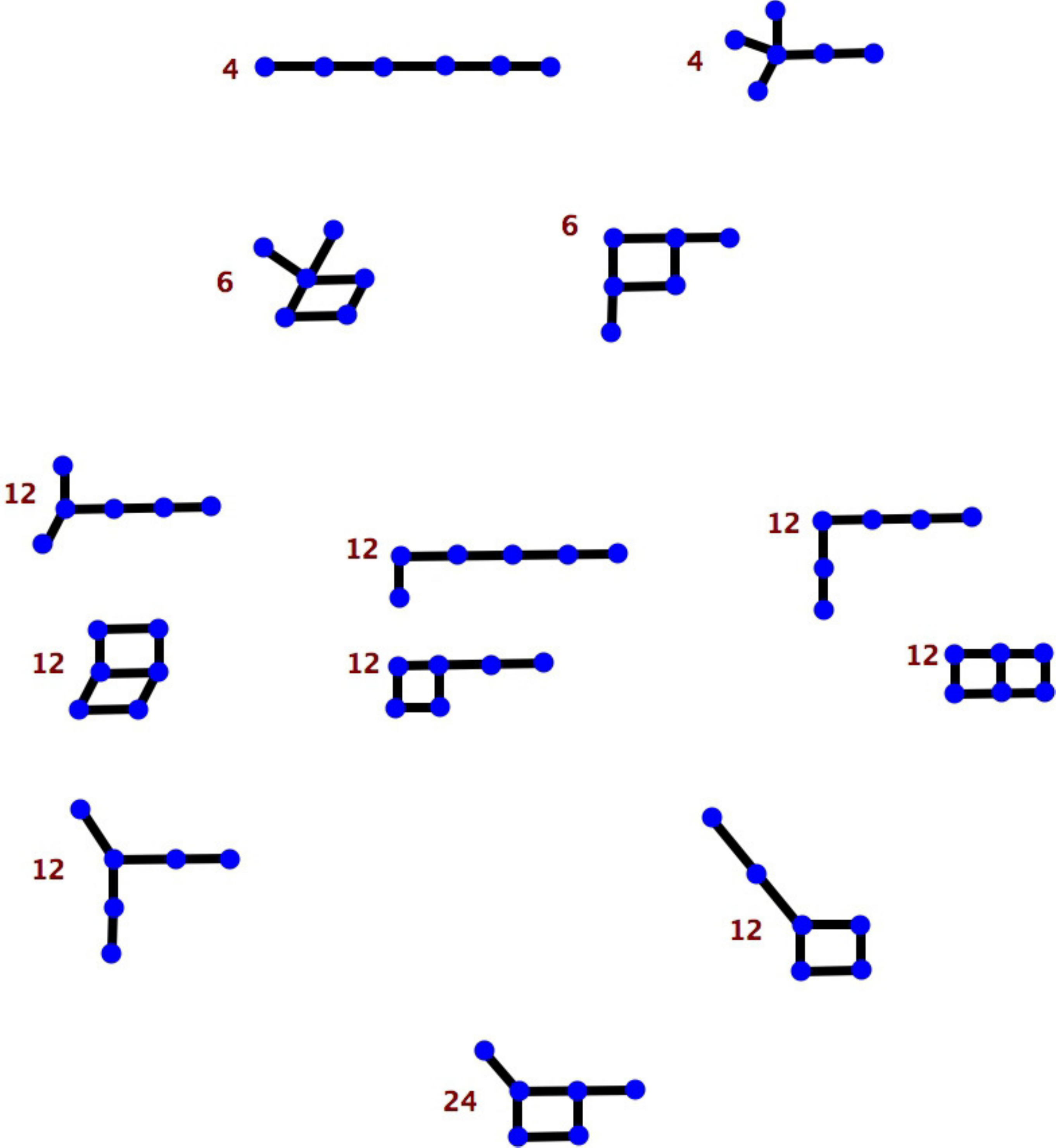}}

\vfill\eject

\section{Statistics of solid partitions}

Since \cite{MM} new ways of enumerating partitions came to the attention of mathematicians and physicists. For example, the ordinary, $2$-dimensional partitions $\lambda$ are in one-to-one correspondence with the irreducible representations  ${\CalR}_{\lambda}$ of the symmetric group $S(|{\lambda}|)$. The space of representations has a natural Plancherel measure, 
\beq
{\mu}_{\lambda} = \frac{1}{|{\lambda}|!} \left( {\rm dim}{\CalR}_{\lambda} \right)^{2}
\label{eq:planch}
\eeq
which was studied for large $|{\lambda}|$ in \cite{VK}. The measure \eqref{eq:planch}, in turn, has a deep three-parametric generalization, interpolating between the uniform (constant) and Plancherel measure. The generalization came from the studies of supersymmetric gauge theories, four-manifold invariants, and representation theory of infinite-dimensional algebras:
\beq
{\mu}_{\lambda}(q_{1}, q_{2}, q_{3}) = \prod_{{\square}\in {\lambda}} \frac{\left( 1- q_{3} q_{1}^{a_{\square}+1} q_{2}^{-l_{\square}} \right)\left( 1- q_{3} q_{1}^{-a_{\square}} q_{2}^{l_{\square}+1} \right)}{q_{3} \, \left( 1- q_{1}^{a_{\square}+1} q_{2}^{-l_{\square}} \right)\left( 1- q_{1}^{-a_{\square}} q_{2}^{l_{\square}+1} \right)} 
\label{eq:qqmeas}
\eeq
where for ${\square} = (i,j) \in {\lambda}$ one defines the arm-length $a_{\square} = {\lambda}_{i} - j$ and the leg-length $l_{\square} = {\lambda}^{t}_{j} - i$ (see \cite{NaG, NaG1, N2, NO} for more explanations and more notations).  
The strange-looking formula \eqref{eq:qqmeas} and its scarier versions for the ensembles of colored partitions \cite{N2, NO} can be pacified by the plethystic exponent presentation (we already used such a presentation in writing the Eq. \eqref{eq:zdq}):
\beq
{\mu}_{\lambda}(q_{1}, q_{2}, q_{3}) = q_{3}^{-| {\lambda} |} \, {\sl E}\left[ (1-q_{3}) \left( N K^{*} + N^{*} K - (1-q_{1})(1-q_{2}) K K^{*} \right) \right]
\eeq
where
\beq
\mathsf{E}\left[  f (x_1, x_2, \ldots , x_p) \right] = {\exp} \, \sum_{l=1}^{\infty} \frac{1}{l} f (x_1^{l}, x_2^{l}, \ldots , x_p^{l})
\label{eq:plexp}
\eeq
and for \eqref{eq:qqmeas} $N = 1$, 
\beq
K = \sum_{(i,j) \in \lambda} q_{1}^{i-1} q_{2}^{j-1}
\eeq
and $f^{*} (x_1, x_2, \ldots , x_p) = f (x_1^{-1}, x_2^{-1}, \ldots , x_{p}^{-1})$. 
The conjecture \cite{NJ} (proven using the results of \cite{Carlsson:2013jka}) 
\beq
\sum_{\lambda} \, {\mu}_{\lambda}(q_{1}, q_{2}, q_{3}) \, {\qe}^{|{\lambda} |} = 
\mathsf{E} \, \left[ \frac{(1-q_{1}q_{3})(1-q_{2}q_{3})}{(1-q_{1})(1-q_{2})q_{3}} \frac{\qe}{1-{\qe}} \right]
\label{eq:5to6}
\eeq
relates the $U(1)$ noncommutative gauge theory in $4+1$ dimensions to the
$(2,0)$ tensor multiplet in $5+1$ dimensions, compactified on an elliptic curve. The fugacity $\qe$ in this correspondence becomes the nome of the elliptic curve ${\qe} = {\exp}\, 2\pi \ii\tau$. The relations \eqref{eq:5to6} and their more sophisticated analogues motivated the general BPS/CFT correspondence conjecture of 
\cite{N1}, and its much more detailed AGT version \cite{AGT}. 

\subsection{The measure}

We start by presenting the explicit formula for the complex-valued measure on the set of solid partitions, which originates in gauge theory. 

Let $q_{a}$, $a = 1, 2, 3, 4$ be non-zero complex numbers, obeying
\beq
\prod\limits_{a=1}^{4}\, q_{a} = 1
\label{eq:tor4}
\eeq
Fix $\mu \in {\BC}^{\times}$, and define, for the indeterminates $x_{1}, \ldots , x_{k} \in {\BC}^{\times}$, the rational function (a symmetric function of $x_{1}, \ldots , x_{k}$):
\begin{multline}
{\CalX}_{k} = \frac{1}{k!} \left( \frac{ q_{4} (1-q_{12})(1-q_{13}) (1-q_{23})}{\sqrt{\mu} (1-q_{1})(1-q_{2})(1-q_{3})(1-q_{4})} \right)^{k} \times \\ \times
\prod_{1 \leq i \neq j \leq k} \frac{\left( x_{j} - x_{i}  \right)  \left( x_{j} - q_{12} x_{i}   \right) \left( x_{j} - q_{23} x_{i}  \right) \left( x_{j} - q_{13} x_{i}  \right)}{\left( x_{j} - q_{1} x_{i}  \right)  \left( x_{j} - q_{2} x_{i}   \right) \left( x_{j} - q_{3} x_{i}   \right) \left( x_{j} - q_{4} x_{i} \right)} \times \\
\times \prod_{i=1}^{k} \frac{1- {\mu} x_{i}}{1-x_{i}} \end{multline}
For the four dimensional (solid) partition ${\rho} \subset {\BZ}_{+}^{4}$ of size $|{\rho}| = k$
define:
\beq
{\sf x}_{\rho} = \left( q_{1}^{a-1}q_{2}^{b-1}q_{3}^{c-1}q_{4}^{d-1} \right)_{(a,b,c,d) \in {\rho}}
\eeq
\beq
{\mathfrak M}({\rho}) =  \sum_{{\sigma} \in S(k)} {\rm Res}_{( x_{{\sigma}(i)} )_{i=1}^{k}  = {\sf x}_{\rho}} \, {\CalX}_{k} \, \prod_{i=1}^{k} \frac{dx_{i}}{x_{i}}
\label{eq:mmeas}
\eeq

\section{Gauge theory}

We now pass to the derivation of \eqref{eq:mmeas}. 
 
\subsection{ADHM\ construction\ in\ four\ complex\ dimensions}

Let $N, K, T$ be complex Hermitian vector spaces, of dimensions $n, k$, and $4$, respectively. 
We assume, in addition, that $T$ is endowed with the fixed holomorphic $4$-form ${\Omega} \in {\Lambda}^{4}T^{*}$, which is compatible with the Hermitian structure, 
\beq
{\Omega} \wedge {\bar\Omega} = {\rm vol} \ . 
\eeq
The space ${\Lambda}^{2}T^{*}$ of exterior two-forms 
has a real structure. Let ${\CalU} \approx {\BR}^{6}$ be the corresponding
real space, so that ${\CalU} \otimes {\BC} = {\Lambda}^{2}T^{*}$.

The symmetry $\Hf$ of the problem is the product
\beq
{\Hf} = U(N) \times U(K) \times SU(4)
\label{eq:symm}
\eeq
The ADHM data consists of the quadruple of matrices $B_{a} \in End(K)$, $a = 1, 2, 3, 4$, which we combine into a linear map:
\beq
{\BB} : K \to K \otimes T
\eeq
and a homomorphism: $I: N \to K$. The commutator
\beq
[{\BB}, {\BB}]: K \to K \otimes {\Lambda}^{2}T
\eeq
can be projected into the self-dual and the anti-self-dual parts. Specifically, 
we define the following $\Hf$-equivariant equations:
\beq
s_{ab} \equiv [B_{a}, B_{b}] + \frac 12 {\Omega}_{abcd} [B_{d}^{\dagger}, B_{c}^{\dagger}] = 0, \qquad  1 \leq a < b \leq 4 \label{eq:sab}
\eeq
and
\beq
{\mu} \equiv \sum\limits_{a=1}^{4} [B_{a}, B_{a}^{\dagger}] + I I^{\dagger}  = {\zeta} \cdot {\bf 1}_{K}
\label{eq:rmm}
\eeq
The equations \eqref{eq:sab} actually imply the stronger equations 
\beq
[B_{a}, B_{b}] = 0\, ,
\label{eq:bab}
\eeq since
\beq
\sum_{1 \leq a< b \leq 4} {\Tr} s_{ab}s_{ab}^{\dagger} =  \sum_{1 \leq a < b \leq 4} {\Tr} [B_{a}, B_{b}][B_{a}, B_{b}]^{\dagger}
\label{eq:ssqab}
\eeq
The equation \eqref{eq:rmm} with $\zeta > 0$ is equivalent to the stability condition:
\beq
{\rm Any\ subspace} \ K' \subset K \ {\rm s.t.}\ I(N) \subset K' 
, \ {\rm and} \ B_{a} (K') \subset K'
\, , \ a =1, 2, 3, 4 \ \Longrightarrow \ K' = K
\label{eq:stabk}
\eeq
The proof is identical to that for crossed instantons \cite{N3}. In fact, it was by studying the crossed and spiked instantons \cite{N3} that we arrived at the equations above. 

The space of commuting quadruples $\BB$ obeying the stability condition \eqref{eq:stabk}
is the celebrated Hilbert scheme of points $Hilb^{[k]}({\BC}^{4})$ on ${\BC}^{4}$. Although the scheme is defined algebro-geometrically over $\BC$, our way of defining it uses the real structure. The advantage of our approach (as we advocated earlier in \cite{MNOP} and later in \cite{N3}) is the possibility of using the conventional Mathai-Quillen representative
for the integrals over what is now called the virtual fundamental  cycle \cite{GP}. The alternative approaches use the perfect obstruction theory \cite{GP}, derived differential geometry \cite{BJ} and other sophisticated techniques whose physical meaning is yet to be clarified. 

By writing $B_{a} = X_{2a-1} + {\ii} X_{2a}$, $a = 1, \ldots 4$, with Hermitian $X_{m}$, $m = 1, \ldots , 8$ we can view the collection of the matrices ${\BB}, {\BB}^{\dagger}$ as the linear
map
\beq
{\bX} : K \to K \otimes {\CalV}
\eeq 
where ${\CalV} \approx {\BR}^{8}$, ${\CalV} \otimes {\BC} = T \oplus T^{*}$. 

\subsubsection{Deformations and obstructions}

We mentioned the obstruction theory above. In our context, each solution of the equations
\eqref{eq:sab}, \eqref{eq:rmm} defines two vector spaces:
\beq
T^{def} = {\rm ker} {\CalD}\, , \qquad T^{obs} = {\rm coker} {\CalD}
\eeq
where ${\CalD}$ is the linearization of the equations and projector onto the subspace orthogonal to the tangent space to the $U(K)$-orbit. The space $T^{def}$ actually has the complex structure (for the same reason the Eqs. \eqref{eq:sab} imply the holomorphic
equations \eqref{eq:bab}. Therefore $T^{def}$ has a canonical orientation. 
The space $T^{obs}$ is only a real vector space. The subtle and important feature of $T^{obs}$ is that it also has an orientation. Even though the dimensions of $T^{def}$
and $T^{obs}$ may jump over different loci in $Hilb^{[k]}({\BC}^{4})$, the orientation ${\rm det}T^{obs}$ stays constant. We show this using the

\subsection{Cohomological field theory}

Let us now explain the origin of our measure \eqref{eq:mmeas}. We will be integrating over the supermanifold (in fact, the supervector space) 
\beq
{\CalX}_{K} = \left( {\Pi} T \left( {\rm Hom}(K, K \otimes {\CalC}) \oplus {\rm Hom}(N,K) \right) \oplus {\rm Lie}U(K) \right)/U(K)
\label{eq:svec}
\eeq
where
\beq
{\CalC} = {\BR}^{2} \oplus {\CalV} \oplus {\CalU} 
\eeq
has real dimension $16+1$. The bosonic variables (fields) of our integration problem are:
$X_{m}, {\sigma}, {\bar\sigma}, H_{ab} = {\sigma}_{ab}^{i} h_{i} = {\ve}_{abcd}H_{dc}^{\dagger}$, $h$  
valued in ${\rm End}(K)$, with $X_{m} = X_{m}^{\dagger}$, 
$m= 1, \ldots , 8$,  
${\sigma}^{\dagger} = {\bar\sigma}$, and seven ``auxiliary'' fields 
$h_{i} = h_{i}^{\dagger}$, $i = 1, \ldots, 6$ and $h = h^{\dagger}$. 
In addition, our bosonic variables include $I \in {\rm Hom}(N, K), 
I^{\dagger} \in {\rm Hom}(K, N)$. The fermionic
variables are
${\Psi}_{m}, {\eta}, {\chi}_{ab} = {\sigma}_{ab}^{i} {\chi}_{i} = {\ve}_{abcd}{\chi}_{dc}^{\dagger}$, and $\chi$, all valued in ${\Pi}{\rm End}(K)$, and
${\psi} \in {\Pi}{\rm Hom}(N,K), {\psi}^{\dagger} \in {\Pi}{\rm Hom}(K,N)$. 
The most important fact about the space \eqref{eq:svec} is the nilpotent
(on the quotient by $U(K)$) odd vector field, which acts on our variables
as follows:
\beq
\begin{aligned}
& {\delta}X_{m} = {\Psi}_{m}, \qquad {\delta}{\Psi}_{m} = [ {\sigma}, X_{m} ], \\
& {\delta}I = {\psi}, \qquad {\delta}{\psi} =  {\sigma} I,\\
& {\delta}I^{\dagger} = {\psi}^{\dagger}, \qquad {\delta}{\psi}^{\dagger} =  
- I^{\dagger} {\sigma}, \\
& {\delta}{\bar\sigma} = {\eta}, \qquad {\delta}{\eta} = [ {\sigma}, {\bar\sigma} ], \\
& {\delta}{\chi}_{i} = h_{i}, \qquad {\delta}h_{i} = [ {\sigma}, {\chi}_{i} ], \\
& {\delta}{\chi} = h, \qquad {\delta}h = [ {\sigma}, {\chi} ] ,\\
& {\delta}{\sigma} = 0 \\
\end{aligned}
\label{eq:cohsusy}
\eeq
The operator $\delta$ is nothing but the equivariant de Rham differential. Its square
${\delta}^{2} = {\CalL}_{\sigma}$ is the infinitesimal gauge transformation
generated by $\sigma$. 

The supersymmetric measure on the space of the eight-dimensional
ADHM data which represents the {\it bulk contribution to Witten index} a la 
\cite{Sav, Smilga} is given by:
\beq
\left\{ \frac{D[bosons]D[fermions]}{{\rm Vol}(U(K))} \right\} \ e^{-{\delta}{\bf\Psi}}
\label{eq:bulk1}
\eeq
where
the measure $D[bosons]D[fermions]$ is the canonical Berezin measure up to a factor $D{\sigma}$
which is uniquely defined (as in  \cite{W}) by normalizing it so as to produce
the Haar measure on $U(K)$ of volume $1$. The {\it gauge fermion}
${\bf\Psi}$ is taken to be equal to:
\beq
{\bf\Psi} = {\rm Re} \left\{ 
\sum_{a<b} {\Tr} \left( {\chi}_{ab}^{\dagger} \left( s_{ab} - H_{ab} \right) + c.c. \right) + \sum_{m} {\Tr} {\Psi}_{m}  [ {\bar\sigma}, X_{m} ]  + {\Tr} 
\left( {\psi}^{\dagger} ({\bar\sigma}I) + c.c. \right) +
{\Tr} {\eta} [{\sigma}, {\bar\sigma}] \right\}
\eeq

\subsubsection{$\Omega$-deformation}

Now let us use the $SU(4)$-symmetry of the equations \eqref{eq:sab}, \eqref{eq:rmm}:
\beq
B_{a} \mapsto U_{a}^{b} B_{b} \, ,
\eeq
with $UU^{\dagger} = 1$, ${\Det}(U) = 1$. One promotes the differential 
\eqref{eq:cohsusy} to the $SU(4) \times U(K)$ equivariant differential ${\delta}_{\ve}$, with important
modifications:
\beq
\begin{aligned}
& 
{\delta}_{\ve}\left( {\Psi}_{2a-1} + {\ii} {\Psi}_{2a} \right) = [ {\sigma}, B_{a}] + {\ve}_{a} B_{a} \, , \\
& {\delta}_{\ve} H_{ab} = [ {\sigma} , {\chi}_{ab} ] + ( {\ve}_{a} + {\ve}_{b} ) {\chi}_{ab} \, , \\
\end{aligned}
\label{eq:eqved}
\eeq
with ${\delta}_{\ve}^{2} = {\CalL}_{\sigma} + {\CalL}_{\ve}$ where
${\rm diag}({\ve}_{1}, {\ve}_{2}, {\ve}_{3} , {\ve}_{4})$ an element of the complexified
Cartan subalgebra of $SU(4)$. 

{} We can now address the choices of the orientation we discussed earlier. They amount to the ordering of the $\chi_{i}$'s in the measure $D{\chi}$. Once this order is fixed the orientation
is chosen globally. In writing \eqref{eq:mmeas} we chose the ordering $D{\chi}_{12} D{\chi}_{13} D{\chi}_{23} D {\chi}_{34} D{\chi}_{42} D{\chi}_{41}$.

\subsubsection{Matter bundle}

In addition to the fermionic symmetry $\delta$ the system $({\BB}, {\BB}^{\dagger}, \ldots )$
has the so-called ghost number $U(1)$ symmetry, under which the
bosonic variables $X_{m}, I, I^{\dagger}, h_{i}, h$, have degree zero, the 
boson $\sigma$ has degree $+2$, the boson $\bar\sigma$ has degree $-2$,
the fermions ${\Psi}_{m}, {\psi}, {\psi}^{\dagger}$ have degree $+1$, 
the fermions ${\chi}_{i}, \chi, \eta$ have degree $-1$. The total
charge of the measure 
\beq
\left\{ \frac{D[bosons]D[fermions]}{{\rm Vol}(U(K))} \right\}
\eeq
also known as the ghost number anomaly or the virtual dimension of the
 moduli space is equal to $2k$.

 In the absence of $\Omega$-deformation the integral \eqref{eq:bulk1} vanishes
 unless we insert some observable of positive ghost number. One possibility, 
 as in \cite{W} is to use the equivariant symplectic form
 \beq
 {\varpi} = {\Tr} \sum_{a=1}^{4} \, {\Psi}_{2a-1} {\Psi}_{2a}  + {\psi} {\psi}^{\dagger}
+ {\Tr} {\sigma} {\mu}
\eeq
which obeys ${\delta}{\varpi}  = 0$, so that \eqref{eq:bulk1} becomes
\beq
Z_{k}^{\zeta} = \int_{{\CalX}_{k}} \left\{ \frac{D[bosons]D[fermions]}{{\rm Vol}(U(K))} \right\} \ e^{\varpi -{\delta}_{\ve}{\bf\Psi}_{\bar\ve}} 
\label{eq:bulk2}
\eeq
where ${\bf\Psi}_{\bar\ve}  = {\bf\Psi} + \ldots $ is a deformed version of $\bf\Psi$. We don't need
the precise expression. 

As in \cite{N3} it is more advantageous to insert more geometric
observables. 
Recall that the space $K$ is the vector bundle over the moduli space 
${\iM}_{k}$ of solutions to \eqref{eq:sab}, \eqref{eq:rmm} modulo $U(K)$.
The natural observable is the equivariant Euler class of $K$, and more generally its Chern polynomial 
\beq
c(m ;K) = \sum_{i=0}^{k} c_{k-i}(K) m^{i}\ .
\label{eq:cmk}
\eeq
By analogy with the
four dimensional instanton calculus \cite{N2} we call $K$ the {\it matter bundle}. The 
operator \eqref{eq:cmk} can be represented using an auxiliary system of bosons ${\tilde H}$ and fermions $\Upsilon$, valued in $K$, and their conjugates ${\tilde H}^{\dagger}$, ${\Upsilon}^{\dagger}$, with the $\delta_\ve$ symmetry acting as:
\beq
{\delta}_{\ve} {\Upsilon} = {\tilde H} \, , \qquad {\delta}_{\ve} {\tilde H} = \left( {\sigma} + m \right) {\Upsilon} \, , \qquad {\delta}_{\ve} {\Upsilon}^{\dagger} = {\tilde H}^{\dagger} \, , \qquad {\delta}_{\ve} {\tilde H}^{\dagger} = - {\Upsilon}^{\dagger} \left( {\sigma} + m \right) 
\eeq
We modify the gauge fermion to 
\beq
{\bf\Psi} \longrightarrow {\bf\Psi} + {\Tr} {\Upsilon}^{\dagger}{\tilde H} 
\label{eq:psim}
\eeq
{}The $\Omega$-deformation violates the ghost number symmetry, in that the
parameters ${\ve}_{a}$ should be assigned the charge $+2$ in order for
the $\delta_{\ve}$ operator to have the same charge $+1$ as the operator
$\delta$. So the $\Omega$-deformed theory has nontrivial partition function both with and without the modification \eqref{eq:psim}. We recover the theory without matter deformation by taking the limit $m \to \infty$, at the same time tuning the fugacity $\qe \to 0$ so that
$m{\qe} = {\Lambda}$ stays finite. This is analogous to the flow from the ${\CalN}=2^{*}$
theory in four dimensions to the pure ${\CalN}=2$ theory. 

The mass parameter $m$ is similar to the equivariant parameters $\ve_{a}$, in that it corresponds to a $U(1)$ symmetry of the vector bundle $K$. If the parameter $m$ is a integral linear combination 
\beq
m \to \sum_{a=1}^{4}  n_{a} {\ve}_{a}\, , \qquad n_{a} \geq 0
\eeq
then the bundle has an equivariant section 
\beq
{\tilde s} = \prod_{a=1}^{4} B_{a}^{n_{a}} \, I
\eeq
which can be used to modify the gauge fermion further to
\beq
{\bf\Psi} \longrightarrow {\bf\Psi} + {\Tr} {\Upsilon}^{\dagger}\left( {\tilde H} - {\tilde s} \right) 
\label{eq:psim2}
\eeq
which by scaling $\tilde s$ can be made the integral localize onto the locus 
\beq
{\tilde s} = 0
\label{eq:locuss}
\eeq
which is the usual representative of the Euler class of $K$. 

\subsubsection{Open string theory}

We expect the fields $I, I^{\dagger}$, ${\Upsilon}, {\Upsilon}^{\dagger}$ result in the quantization of open strings connecting $D0$-branes and $D8$ and anti-$D8$ branes, respectively \cite{NPrF}. The $X_{m}$ etc. are the usual $0-0$ strings. 

The deformation \eqref{eq:psim2} and the subsequent localization to \eqref{eq:locuss}
surely reflects an interesting spacetime event. For example, when  $m= {\ve}_{a}$ for $a = 1, 2, 3$, or $4$ our model, we believe, describes the effects of the tachyon condensation in the $D8$-$\overline{D8}$ system, resulting in their annihilation leaving behind a single $D6$ brane stretched in the ${\BC}^{3}_{{1\ldots}\check{\tiny a}{\ldots 4}} \times {\BR}^{1}_{0}$ direction (we use the notations of \cite{N3, NPr}). For the discussion of the tachyon condensation in the context of string field theory and boundary string field theory see
\cite{Sen1, Sen2,  BSZ, GS, GS1, Akhmedov, WittenK}. 

\subsubsection{Quantum mechanics}

Finally, the measure \eqref{eq:mmeas} comes from the $K$-theoretic, or loop space analogue of 
\eqref{eq:bulk1}. It is the localization computation of the path integral representing the Witten index:
\beq
{\CalZ}_{k}^{\rm inst} (q_{a}; {\mu}) = 
{\Tr}_{{\bf H}^{\rm phys}} \, (-1)^{F} U e^{-{\beta}{\hat H}} = 
\int \, \frac{Dg}{{\rm Vol}U(K)} \ {\Tr}_{{\bf H}^{\rm unphys}} \, (-1)^{F} U \, g \, e^{-{\beta}{\hat H}}
\label{eq:bulk3}
\eeq
where $U = {\rm diag}(q_{1}, q_{2}, q_{3}, q_{4}) \times {\mu}$ imposes the $SU(4) \times U(1)$ twisted
boundary conditions in the supersymmetric quantum mechanics of our fellow friends
${\BB}, {\BB}^{\dagger}, \ldots, {\Upsilon}, {\Upsilon}^{\dagger}$. The second  
expression in  \eqref{eq:bulk3} corresponds to imposing the $U(K)$ Gauss law by averaging
over the gauge group. 

Our theory computes the
equivariant index of Dirac operator coupled to the alternating sum of vector bundles:
\beq
{\CalZ}_{k}^{\rm inst} (q_{a}; {\mu}) = \left[  \, \sum_{i} (-{\mu})^{i} \bigwedge^{i} K \, \right]_{K({\iM}_{k})} \sim \int_{\left[ {\iM}_{k} \right]^{\rm virt}} {\bf A} (T_{{\iM}_{k}}) \, \sum_{i} (-{\mu})^{i} {\rm Ch}\left( \bigwedge^{i} K \right)
\label{eq:inddir}
\eeq
where $\mu$ is the additional equivariant weight, and ${\bf A}$ stands for the virtual $A$-roof genus:
\beq
{\bf A}(T_{X}^{def}-T_{X}^{obs}) = \prod_{i} \frac{x_{i}}{2 sinh(x_{i}/2)} \, \left( \prod_{j} \frac{y_{j}}{2 sinh(y_{j}/2)} \right)^{-1}
\eeq
with $x_{i}$'s being the Chern roots of the deformations, the $y_{j}$'s the Chern roots of the obstructions.

Our ultimate goal is the generating function
\beq
{\CalZ}^{\rm inst} (q_{a}; {\mu}; {\qe})= \sum_{k=0}^{\infty} {\qe}^{k} {\CalZ}_{k}^{\rm inst}
\eeq
By playing with the coupling constants (using the $\delta_{\ve}$-invariance of the measure) 
one arrives at the contour integral expression which leads to \eqref{eq:mmeas}. This is 
analogous to the manipulations in \cite{MNS, MNS2, N2}. 

\subsection{Enter solid partitions}

The residues in the contour integral are the fixed points of the $U(1)^3$ action on $Hilb^{[k]}({\BC}^{4})$, i.e. the zeroes of $\delta_{\ve}$. We claim these are the solid partitions 
$\rho$ of size $|{\rho}| = k$.

The solid partition $\rho$ can be described in several ways: 
as a monomial ideal ${\CalI}_{\rho} \subset {\BC} [z_{1}, z_{2}, z_{3}, z_{4}]$ in the ring of polynomials in four variables such that the quotient
\beq
K = {\BC} [z_{1}, z_{2}, z_{3}, z_{4}] / {\CalI}_{\rho}
\eeq
has finite dimension. Recall that the vector subspace ${\CalI} \subset {\BC} [z_{1}, z_{2}, z_{3}, z_{4}]$ is called an ideal if for any $v \in {\CalI}$ and $f \in {\BC} [z_{1}, z_{2}, z_{3}, z_{4}]$, $f v \in {\CalI}$. The polynomials $g_{1}, \ldots , g_{r}$ are called the generators of an ideal ${\CalI}$
if for any $v \in {\CalI}$ there exist polynomials $f_{1}, \ldots , f_{r}$, such that
\beq
v = f_{1}g_{1} + f_{2}g_{2} + \ldots + f_{r}g_{r}
\eeq
Finally, the ideal ${\CalI}_{\rho}$ is called {\it monomial} if it has a basis of monomial generators. 
The four dimensional torus ${\BT}_{4} = ({\BC}^{\times})^{4}$ acts on ${\BC} [z_{1}, z_{2}, z_{3}, z_{4}], {\CalI}_{\rho}$, and on $K$:
\beq
f \mapsto f^{q} = (q_{1}, q_{2}, q_{3}, q_{4})\cdot f , \qquad f^{q} (z_{1}, z_{2}, z_{3}, z_{4}) = f(  
q_{1}z_{1}, q_{2}z_{2}, q_{3}z_{3}, q_{4}z_{4})
\eeq
So the ideal is $({\BC}^{\times})^{4}$-invariant iff it is monomial. 

Actually, in our story we don't have the full four dimensional torus $({\BC}^{\times})^{4}$
at our disposal, only the maximal torus of $SU(4)$, i.e. $U(1)^3$. But we can easily prove, as in \cite{N3}, that $(B_{1}, B_{2}, B_{3}, B_{4}, I)$ obeying \eqref{eq:sab}, \eqref{eq:rmm}
and 
\beq
{\ve}_{a} B_{a} + [{\sigma}, B_{a} ] = 0 \, , \qquad a = 1, \ldots , 4
{\sigma} I = 0
\label{eq:ffp1}
\eeq
with generic $\ve_a$'s obeying $\sum_a \ve_a = 0$ actually obeys \eqref{eq:ffp1} with arbitrary quadruple $({\ve}_a)_{a=1}^{4}$, i.e. the generator of the four dimensional torus
$({\BC}^{\times})^{4}$. To avoid confusion, the compensating infinitesimal transformation $\sigma$ in \eqref{eq:ffp1} depends on $\ve_a$'s. The idea in \cite{N3} is to note that the
matrix $N = B_{1}B_{2}B_{3}B_{4}$ commutes with the $U(1)^{3}$-action, is nilpotent, so by Jacobson-Morozov theorem includes in the $sl_2$-triple, also commuting with the $U(1)^3$ -action. The Cartan generator of this $sl_2$ furnishes the fourth torus generator. 

One can show that the $T^{def}$ space is a representation of this torus $({\BC}^{\times})^{4}$. The obstruction $T^{obs}$ is not though.

\section{Instanton partition function: the conjecture}

Our conjecture states:
\beq
{\CalZ}^{\rm inst} (q_{1}, q_{2}, q_{3}, q_{4}, ; {\mu}; {\qe}) = {\mathsf{E}} \left[ {\CalF}^{\rm inst} \right] = {\exp} \, \sum_{k=1}^{\infty} \, \frac{1}{k} \, {\CalF}^{\rm inst} (q_{1}^{k}, q_{2}^{k}, q_{3}^{k}, q_{4}^{k},  {\mu}^{k}, {\qe}^{k})
\label{eq:pfn}
\eeq
where
\beq
{\CalF}^{\rm inst} (q_{1}, q_{2}, q_{3}, q_{4}, {\mu}, {\qe}) =  \frac{[q_{1}q_{2}][q_{1}q_{3}][q_{2}q_{3}][{\mu}]}{[q_{1}][q_{2}][q_{3}][q_{4}][\sqrt{\mu} {\qe}][\sqrt{\mu}{\qe}^{-1}]}
\label{eq:freeen}
\eeq
and we used the (unconventional) notation:
\beq
[X] = X^{\frac 12}- X^{-\frac 12}
\label{eq:brack}
\eeq
We have explicitly checked this conjecture up to six instantons (i.e. modulo ${\qe}^7$), so as to not fall on the fate of \cite{MM}. Note that the six-instanton brut-force fit of the signs (the choice of the local orientation of the obstruction space) would require 
\beq
2^{140} \sim 10^{42} 
\eeq
attempts, which would probably take longer than the age of the Universe ($\sim 4 \cdot 10^{17}$ sec) on a regular computer.  Our success is a strong indication the conjecture is correct. 

The Eqs. \eqref{eq:pfn}, \eqref{eq:freeen} have several interesting specifications.

\subsection{Three dimensional DT theory}

Let ${\mu} = q_{a}$, for $a = 1,2, 3$, or $4$. In this case the four dimensional partitions are confined to a three-dimensional subspace $i_{a} = 0, 1$. Our main formula \eqref{eq:freeen}
in this case specifies to (we took $a= 4$ for definiteness) 
\beq
{\CalF}^{\rm inst}_{3} (q_{1}, q_{2}, q_{3}, q_{4}, {\qe}) =  \frac{[q_{1}q_{2}][q_{1}q_{3}][q_{2}q_{3}]}{[q_{1}][q_{2}][q_{3}][\sqrt{q_{4}} {\qe}][\sqrt{q_{4}}{\qe}^{-1}]}
\label{eq:freeen}
\eeq
which, with the addition of the so-called perturbative contribution to the free energy:
\beq
{\CalF}^{\rm pert}_{3} (q_{1}, q_{2}, q_{3})  = \frac{\sqrt{q_{4}} + \sqrt{1/q_{4}}}{[q_{1}][q_{2}][q_{3}]}
\label{eq:pertf}
\eeq
can be cast in the surprisingly $S(5)$-symmetric form
\beq
{\CalF}_{3} \equiv {\CalF}^{\rm pert}_{3}  + {\CalF}^{\rm inst}_{3}  = \frac{\sum\limits_{A=1}^{5} [Q_{A}^{2}]}{\prod\limits_{A=1}^{5} [ Q_{A} ]}
\label{eq:11dsugra}
\eeq
where $Q_{\alpha} = q_{\alpha}$, ${\alpha} = 1, 2, 3$, $Q_{4} = \sqrt{q_{4}}{\qe}$, $Q_{5}= \sqrt{q_{4}}{\qe}^{-1}$, 
\beq
\prod\limits_{A=1}^{5} Q_{A} = 1
\eeq
and the diagonal matrix $U = {\rm diag}(Q_{A}) \in SU(5)$ (in case when $|q_{a}| = |p| = 1$)
is a twist natural for the supersymmetric partition function of $M$-theory \cite{NZ04, NJ},
\beq
Z = {\Tr}_{{\CalH}_{11d}} \, (-1)^{F}  U
\eeq
The conjecture of \cite{NZ04, NJ}
\beq
Z = {\exp} \, \sum_{n=1}^{\infty} \frac{1}{n}  {\CalF}_{3} ( q_{a}^{n}, {\qe}^{n} ) 
\label{eq:pfn11}
\eeq
was recently proven in \cite{AO15}. See also \cite{NO14} for the extension of the theory beyond the points, to account for curves (membranes). 

\subsection{Weaker (cohomological) conjectures}

In the limit $b \to 0$, with $q_{a} = e^{b {\ve}_{a}}$, ${\mu} = e^{bm}$, with ${\qe}, m$ and ${\ve}_{a}$  kept finite, 
our conjecture becomes the statement about the cohomological theory, e.g. a partition function
of the $8$-dimensional super-Yang-Mills theory in the $\Omega$-background:
\beq
{\CalZ}^{\rm inst}_{8} \to {\exp} \, \frac{m({\ve}_{1}+{\ve}_{2})({\ve}_{1}+{\ve}_{3})({\ve}_{2}+{\ve}_{3})}{{\ve}_{1}{\ve}_{2}{\ve}_{3}{\ve}_{4}} \, \sum_{n=1}^{\infty} \frac{1}{n}  \frac{{\qe}^{n}}{(1-{\qe}^{n})^{2}} = {\CalM}_{3}({\qe})^{\frac{m({\ve}_{1}+{\ve}_{2})({\ve}_{1}+{\ve}_{3})({\ve}_{2}+{\ve}_{3})}{{\ve}_{1}{\ve}_{2}{\ve}_{3}{\ve}_{4}}}
\label{eq:mcm8}
\eeq   
In the global situation, where one studies the analogous moduli space of sheaves on a Calabi-Yau four-fold $\CalX$, $c_{1}(T_{\CalX})=0$, as in \cite{BJ, Cao1, Cao2, Cao3}, the exponent in \eqref{eq:mcm8} becomes an integral of the charactertic class:
\beq
\int_{\CalX} c_{3}(T_{\CalX}) c_{1}({\CalL})
\eeq
where $\CalL$ is a line bundle, representing the Chan-Paton gauge bundle on the $\overline{D8}$-brane (so that ${\mu} \sim e^{c_{1}({\CalL})}$). 

As this paper was being prepared for submission we learned that
\eqref{eq:mcm8} was independently checked up to the $\qe^6$ order
(five instantons) in \cite{CaoNew}.

\subsubsection{Singularities and speculations}

As in \cite{N2}, the limit $q_{a} \to 1$ is a thermodynamic limit, in which the partition function \eqref{eq:pfn} behaves as the partition function of a non-ideal gas, confined to the volume $\sim 1/{\rm log}(q_{a}) \to \infty$. This is easy to understand: our instantons are no longer confined to the fixed point $0 \in {\BC}^4$, instead they are free to move along the ${\BC}^{1}_{a}$-plane. 
However, our conjecture implies that even with the $q_{a}$ finite the partition function may 
develop the thermodynamic behavior, namely, when 
\beq
{\qe} \to {\mu}^{\pm \frac 12}
\label{eq:newdim}
\eeq

\section{Conclusions and speculations}

This non-perturbative growth of the additional dimension of our problem (with the tuning \eqref{eq:newdim} of the coupling constant) is of course reminiscient of the emergence of the eleventh dimension in the strong coupling limit of the IIA string \cite{WMth}. What makes our story tantalizing is that it is \eqref{eq:pfn11} which is related to the partition function of the eleven dimensional theory, while our \eqref{eq:pfn} contains more degrees of freedom. If we naively attribute the number of factors in the denominator in \eqref{eq:freeen} to the number of complex spatial dimensions, we are forced to conclude that the theory of solid partitions hints at some thirteen-dimensional theory, which we shall call the $M_{13}$-theory. Is there a room for a topological suspension of F-theory? 
One argument in favor of $M_{13}$ is the existence of $D8$-branes, the nine-dimensional objects. Their geometric realization  could involve the compactification of $M_{13}$ on a Taub-Nut space, just like the $D6$ branes arise from $M$-theory.  

The $S(5)$ symmetry of \eqref{eq:11dsugra}, which is the enhancement of the $S(3) \times S(2)$ symmetry (the permutations of $q_{\alpha}$'s and the nonperturbative symmetry ${\qe} \to {\qe}^{-1}$)  is the Weyl group of $SU(5)$ -- the supersymmetry preserving subgroup of the spatial rotations in the eleven dimensional Poincare group.  The master partition function
\eqref{eq:freeen} has only the $S(4) \times S(2)$ symmetry (the permutations of $q_a$ and the non-perturbative symmetry ${\qe} \to {\qe}^{-1}$). This is the Weyl group of the group 
$SU(4) \times SU(2)$, which is apparently the supersymmetry preserving global symmetry of the mysterious $M_{13}$-theory.

The cohomological limit \eqref{eq:mcm8} of our formula suggests an extension of topological strings. Recall that one of the curious coincidences around the ${\CalM}_{3}({\qe})$ function (first observed by R.~Dijkgraaf) is that it reproduces the all-genus degree zero Gromov-Witten invariants of a Calabi-Yau threefold. In the equivariant form, the all-genus $A$ model partition function of ${\BC}^{3}$ is equal to:
\beq
Z^{\rm top} ({\ve}_{1}, {\ve}_{2}, {\ve}_{3} ; \hbar) = {\exp}\, \sum_{g=0}^{\infty} {\hbar}^{2g-2} {\CalF}_{g} ({\ve}_{1}, {\ve}_{2}, {\ve}_{3}) = {\CalM}_{3}({\qe})^{\frac{({\ve}_{1}+{\ve}_{2})({\ve}_{1}+{\ve}_{3})({\ve}_{2}+{\ve}_{3})}{{\ve}_{1}{\ve}_{2}{\ve}_{3}}}
\eeq
where $\qe = - e^{\ii\hbar}$, $\hbar$ is the closed string coupling constant, the symmetry $\hbar \to - \hbar$ is the consequence of the orientability of the worldsheets in the $A$ model, and (cf. \cite{BCOV}) 
\beq
{\CalF}_{g} ({\ve}_{1}, {\ve}_{2}, {\ve}_{3}) = \int_{\overline{{\CalM}_{g}}({\BC}^{3})} c({\ve}_{1}; {\bH}_{g})c({\ve}_{2}; {\bH}_{g}) c({\ve}_{3}; {\bH}_{g}) =  \frac{({\ve}_{1}+{\ve}_{2})({\ve}_{1}+{\ve}_{3})({\ve}_{2}+{\ve}_{3})}{{\ve}_{1}{\ve}_{2}{\ve}_{3}}  \frac{B_{2g}B_{2g-2}}{2g (2g)! (2g-2)!}
\eeq
using 
\beq
\int_{{\BC}^{3}}^{\ve} 1 = \frac{1}{{\ve}_{1}{\ve}_{2}{\ve}_{3}} 
\eeq
and the theorem proven in \cite{FR}. The correspondence between the perturbative
closed topological string and the Donaldson-Thomas theory (in which the count of plane partitions arises naturally) led to the GW/DT correspondence \cite{INOV, MNOP}. It seems unlikely that our cohomological theory \eqref{eq:mcm8} could be explained by some kind of GW/DT correspondence for the Calabi-Yau fourfolds \cite{KP}. Instead, an extension of the \cite{NO14} theory seems more adequate.

We should also point out a possible connection to the ``ordinary'' 
 three dimensional gravity. As we explain in the appendix, the solid partitions can be visualized as tesselations of the three dimensional Euclidean space, by squashed cubes. In this way we get an ensemble of random three dimensional geometries, analogous, in a way, to the ensembles of two dimensional random geometries described by the matrix models
 of two dimensional gravity \cite{GM, DS, BK, Chekhov:1988nd}. It would be nice to connect our tesselation model to the model of \cite{BMS}.

Our final speculation concerns the algebraic aspects of the moduli spaces ${\iM}_{k}$
and their upgraded versions corresponding to the toric sheaves on toric Calabi-Yau fourfolds with nontrivial ${\rm ch}_{k}$, $k = 2, 3, 4$ (our fat points only have ${\rm ch}_{4} = k$). We expect the equivariant $K$-theory of these spaces act on the cohomological Hall algebras of threefolds, constructed in \cite{KS},  Nakajima algebras \cite{NaG1, Na2, Na3} and the quiver $W$-algebras \cite{KimP} corresponding to surfaces, and quantum toroidal algebras \cite{Feigin}. We call the conjectural superseding algebraic structure the   
{\it Mama}-algebra, although we suspect it is not an algebra in the ordinary sense.

\section{Acknowledgments}

The main formula \eqref{eq:pfn}, \eqref{eq:freeen} has been presented at several venues in 2016-2017: SCGP (Stony Brook, Oct 2016, Jan 2017)\footnote{http://scgp.stonybrook.edu/video_portal/video.php?id=3021}, UC Berkeley (Nov 2016), Landau Institute (Chernogolovka, Dec 2016), HMI Trinity College (Dublin, Feb 2017), CIRM (Luminy, Mar 2017), IAS (Princeton, May 2017), HShE (Moscow, May 2017), LMS regional meeting in Loughborough University (Sept 2017). I thank the organizers and the audience of all these meetings for the inspiration and useful questions. I also thank  M.~Bershtein, S.~Donaldson, I.~Frenkel, H.~Nakajima,  A.~Okounkov, M.~Kontsevich and Y.~Soibelman for interesting discussions. 
Part of the work was done during my 2016, 2017 visits to the IHES (Bures-sur-Yvette, France). I am grateful to this wonderful institution for its hospitality. 

\section{Appendix A. (Hyper)cubes\ and\ their\ projections}

In this appendix we recall some geometry used in visualizing the plane and solid partitions. 

\subsection{Three dimensional cube and its projection}

\hbox{\vbox{\hbox{In drawing the three dimensional} 
\hbox{partitions on the two dimensional sheet of paper}
\hbox{we employ a projection, which makes visible only}
\hbox{half of the faces of the cubes:}} \qquad\includegraphics[width=3cm]{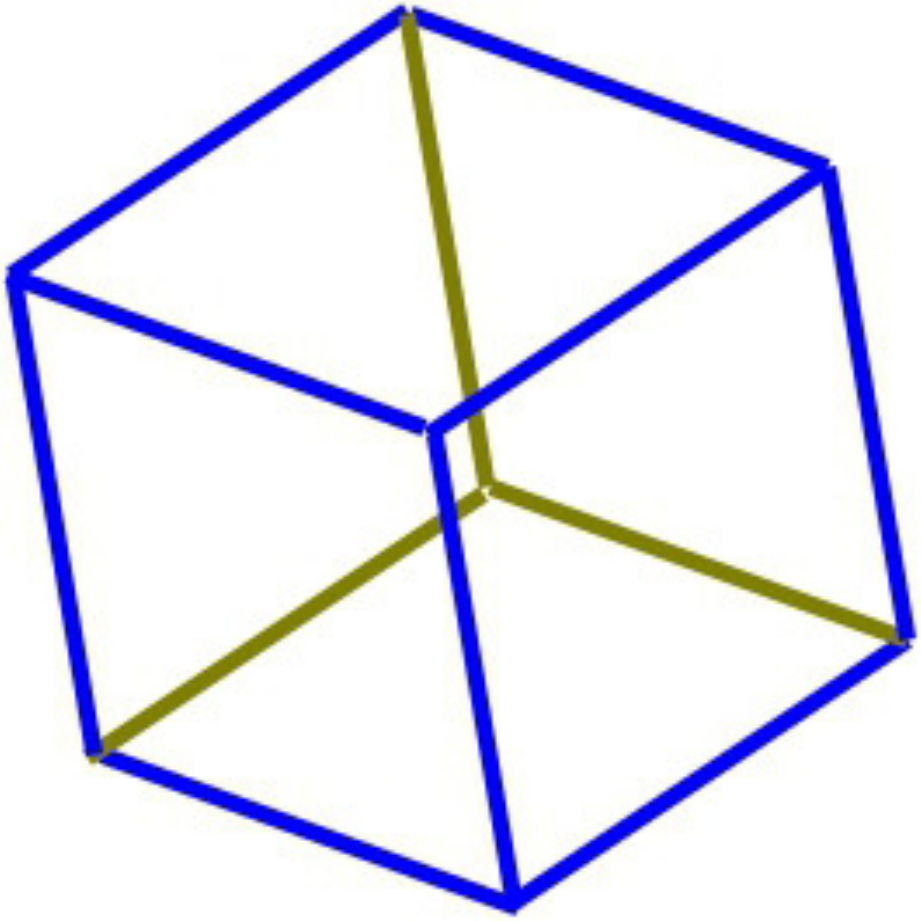}} 

\bigskip

Moreover, out of six faces of a cube only three can be seen on the projection. 
If we choose the $(1,1,1)$ axis as the line of the perspective, then the three faces of a cube land onto three rombi. 

\bigskip
\bigskip\centerline{\includegraphics[width=3cm]{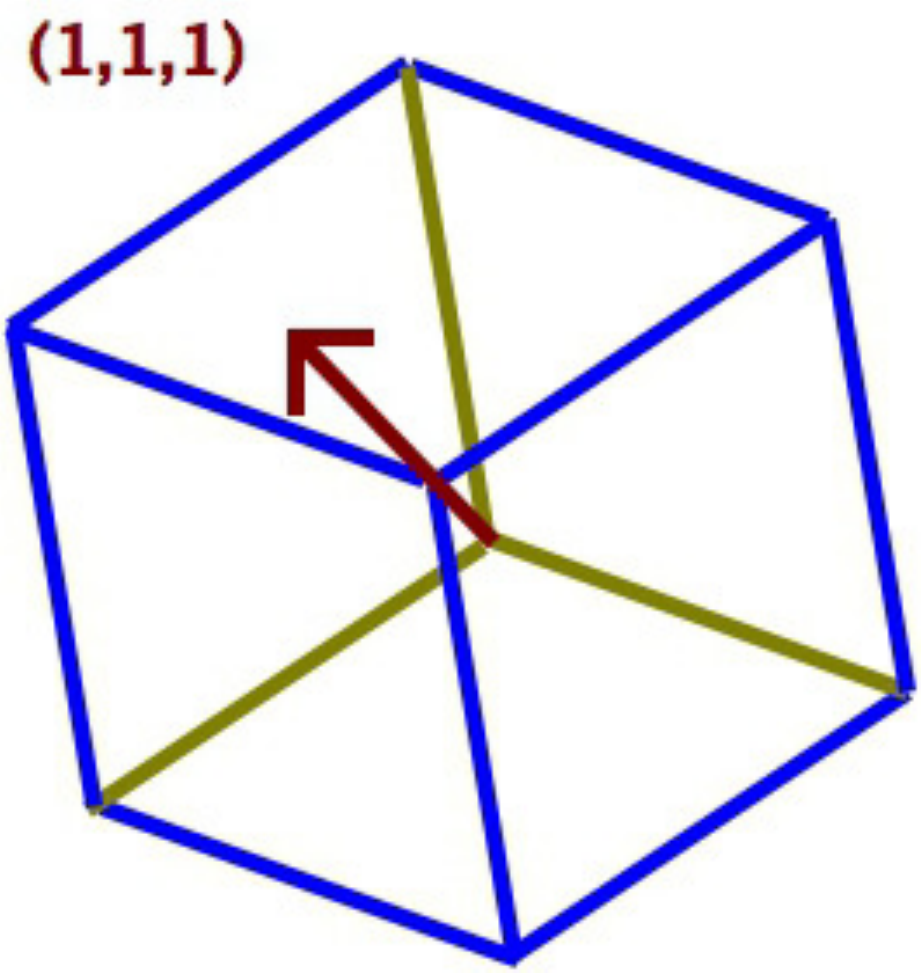}$\Longrightarrow$ \includegraphics[width=3cm]{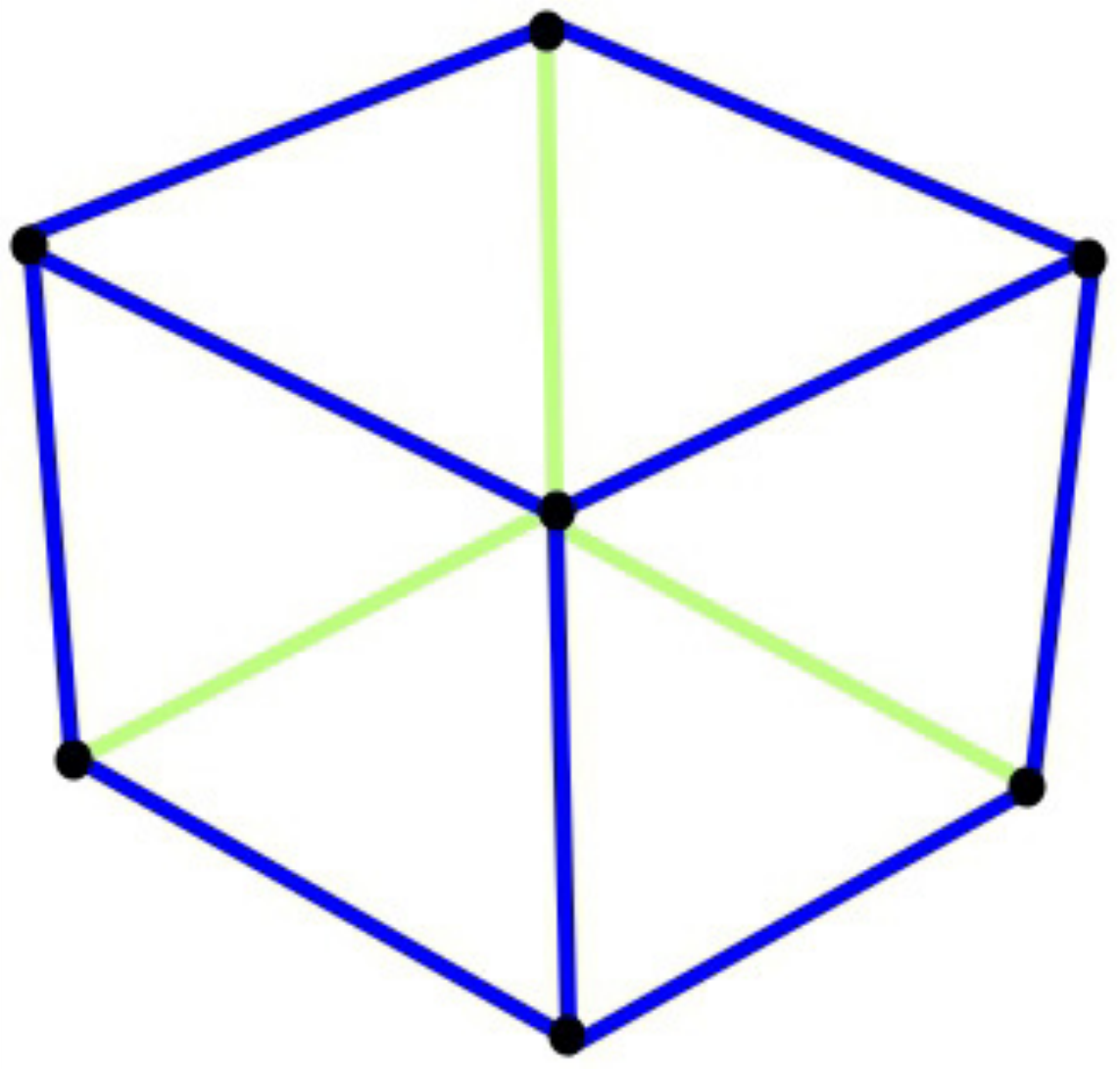}$\Longrightarrow$\includegraphics[width=3cm]{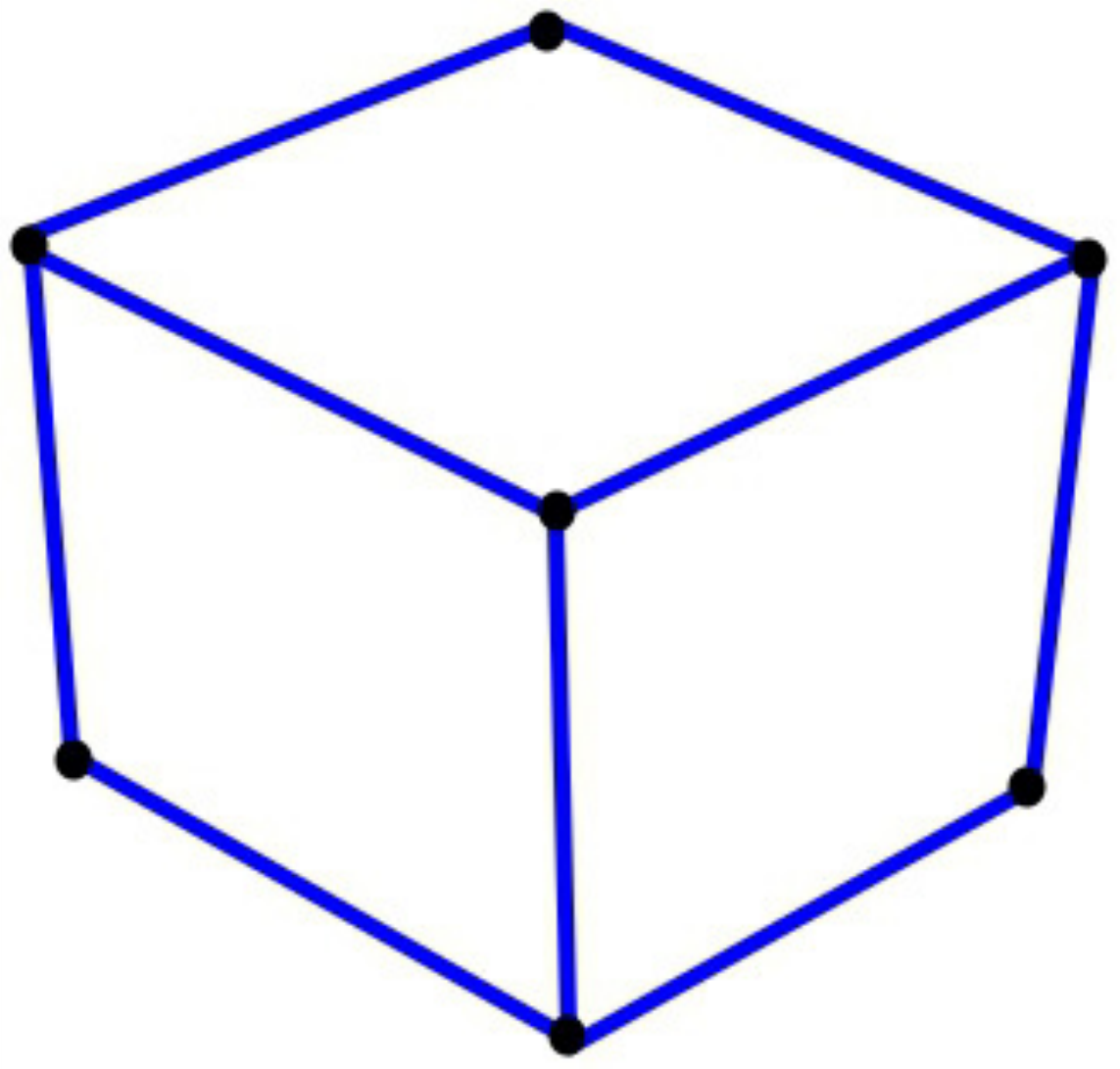}}
\bigskip
\bigskip

Each rombus is characterized by the angles ${\alpha}$ and ${\beta}$ between the
adjacent edges: 
\beq
{\rm cos}({\alpha}) = - \frac 12 \, , \qquad {\beta} = \frac{\pi}{2} - {\alpha}
\eeq
\centerline{\includegraphics[width=5cm]{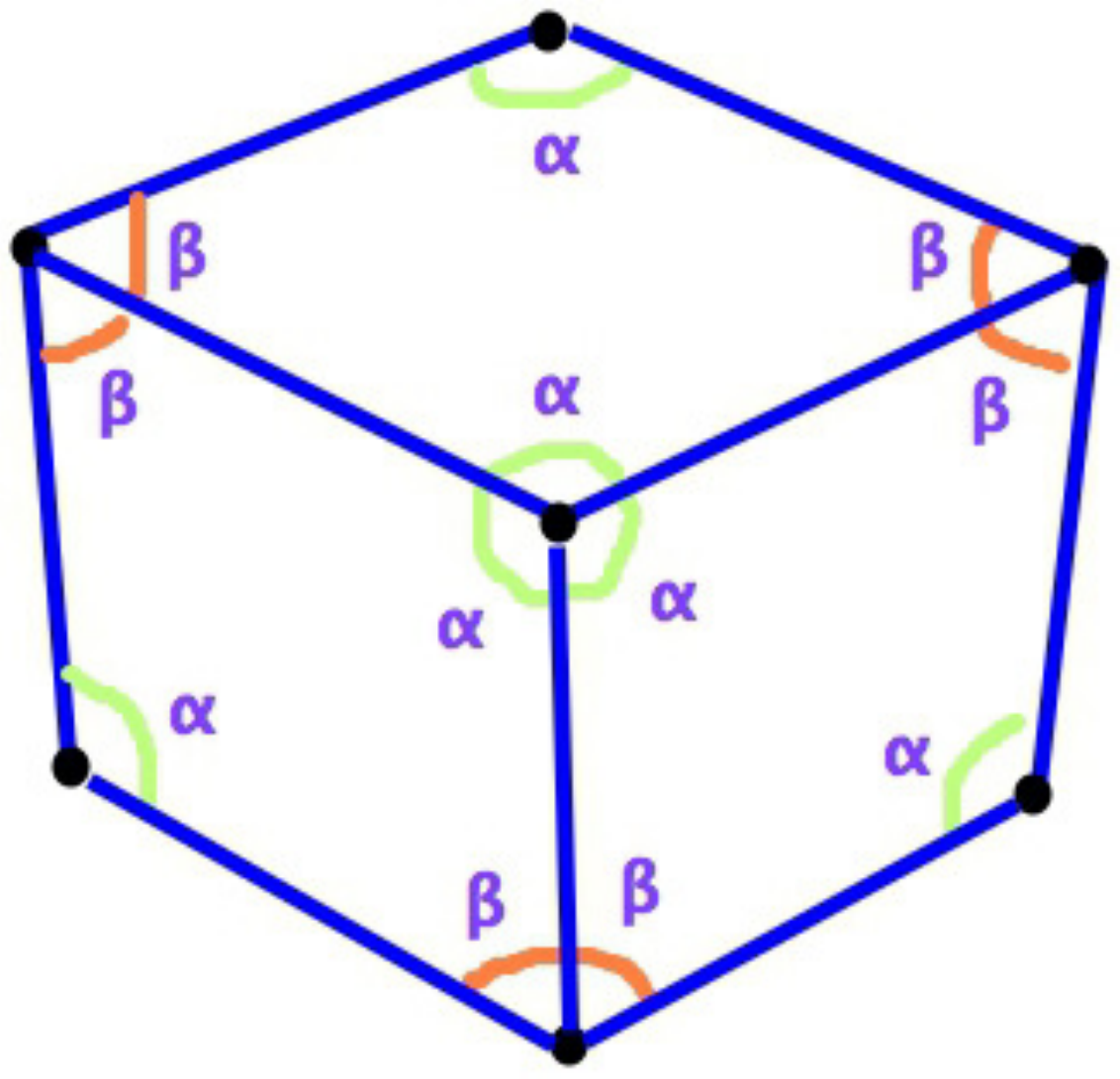}}

{}Using the $(1,1,1)$ projection we map the plane partitions to the dimer configurations on the hexagonal lattice on a plane. See \cite{OR, KOS} for the recent developments in the studies of the thermodynamics of these configurations, and \cite{Fisher} for the early exact solutions.

\subsubsection{Four dimensional hypercube and its projection}

We use the notation $\4 = \{ 1 , 2, 3, 4 \}$ as in \cite{N3}. 
Let ${\be}_{1} = (1, 0,0,0), \ {\be}_{2} = ( 0, 1, 0, 0), \ {\be}_{3} = (0,0,1,0),\ {\be}_{4} = (0,0,0,1)$ be the orthogonal basis in ${\BR}^{4}$. 
Let ${\be} = \frac 12 ( 1, 1,1,1)$. Consider the unit hypercube centered at ${\bn} \in {\BZ}^{4}$: 
\beq
H_{\bn} = \Biggl\{ \, {\br} \ | \ {\br} \in {\BR}^{4}, \ 0 \leq ( {\br}  - {\bn}  , {\be}_{i} ) \leq 1, \ i \in {\4} \ \Biggr\}
\eeq
The boundary of the hypercube is the union of eight cubes: 
\beq
{\partial}H_{\bn} = \bigcup_{i=1}^{4} \ C_{{\bn}, i}^{+} \cup C_{{\bn}, i}^{-} 
\eeq
with $C_{{\bn}, i}^{\pm} = \Biggl\{ \, {\br} \ | \ {\br} \in {\BR}^{4}, \ 0 \leq ( {\br}  - {\bn}  , {\be}_{j} ) \leq 1, \ j \in {\4} \, , \, ( {\br}  - {\bn}  , {\be}_{i} )  = \frac 12 ( 1 \pm 1) \ \Biggr\}$.

Consider the projection $p: {\BR}^{4} \to {\BR}^{3}$ along the ${\be}$-direction:
\beq
p ({\br} ) = {\br}  - ( {\br}, {\be} ) {\be}
\eeq
Let ${\ve}_{i}$, $i \in {\4}$ be the unit vectors in the directions of $p({\be}_{i})$: 
\beq
{\ve}_{i} = \frac{2}{\sqrt{3}} \ p ({\be}_{i}) \, . \eeq
We can choose the orthonormal basis ${\eta}_{1}, {\eta}_{2}, {\eta}_{3}$ in ${\BR}^{3}$, such that
\beq
\begin{aligned}
& {\ve}_{4} = {\eta}_{3}, \\
& {\ve}_{1} = \frac{2\sqrt{2}}{3} {\eta}_{1}  - \frac 13 {\eta}_{3}, \\
& {\ve}_{2} = - \frac{\sqrt{2}}{3} {\eta}_{1} + \frac{\sqrt{2}}{\sqrt{3}} {\eta}_{2}  - \frac 13 {\eta}_{3}, \\
& {\ve}_{3} = - \frac{\sqrt{2}}{3} {\eta}_{1} -\frac{\sqrt{2}}{\sqrt{3}} {\eta}_{2}  - \frac 13 {\eta}_{3} \\
\end{aligned}
\eeq
Then ${\Sigma}_{{\bn}, i}^{\pm} = p(C_{{\bn}, i}^{\pm})$ is a three-dimensional polytope: 
\beq
{\Sigma}_{{\bn}, i}^{\pm} = \Biggl\{ \, {\bx} \, | \, {\bx} \in {\BR}^{3}, \ {\bx} = \sum_{j\in {\4}} t_{j} {\ve}_{j} , \ 
0 \leq t_{j} \leq 1, \ j \in {\4} \, , \, t_{i}  = \frac 12 ( 1 \pm 1) \ \Biggr\}
\eeq

\centerline{\includegraphics[width=5cm]{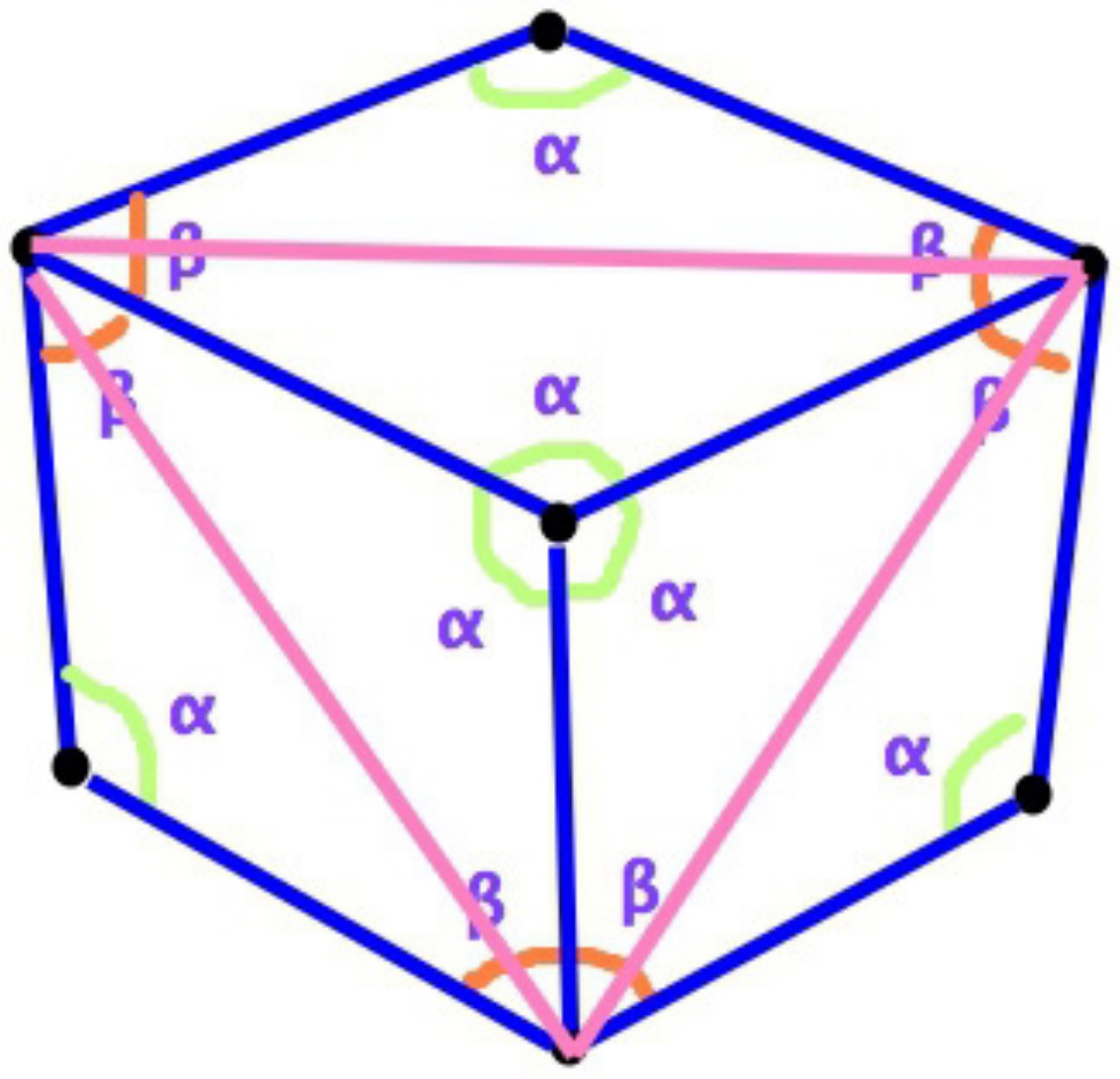}$\Longrightarrow$
\includegraphics[width=5cm]{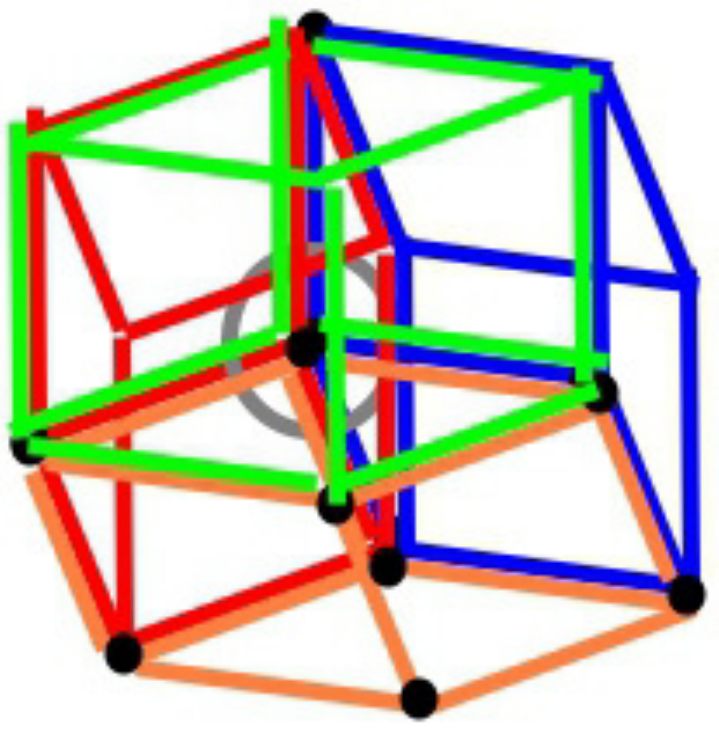}}
On the projection to ${\BR}^{3}$ we see only ${\Sigma}_{{\bn}, i}^{+}$, the four squashed hypercubes. The corresponding angle $\alpha$ is easy to compute from 
\beq
{\rm cos}({\alpha}) = ( {\ve}_{i}, {\ve}_{j} ) = - \frac 13 \,  , \ i \neq j
\label{eq:tetrangle}
\eeq
As the dimer configurations can be drawn on the two dimensional sheet of paper and then printed on a printer, our squashed cubes are best printed on the 3D printer\footnote{An excellent suggestion by A.~Abanov}. We welcome enthusiasts to generate solid partitions and print them. 

\section{Appendix C. Curiosities}

The angle ${\alpha} = {\rm cos}^{-1} \left( - \frac 13 \right)$, which defines the ``squashed cube'' has appeared in the context of supersymmetric field theories in four dimensions
in \cite{Gates1}. The $4+2$ split of the symmetry of the higher ($12+$) dimensional 
(gauge) field theory has been discussed in \cite{Chandrasekar}. Finally, many people have asked me whether the conjecture \eqref{eq:freeen} can be used to correct \cite{MM}. To this end one would need to come up with the limit of the parameters $q_a$'s etc such that the measure \eqref{eq:mmeas} becomes uniform. The numerical studies \cite{BGPr} suggest this is unlikely. Our conclusion is the uniform measure on solid partitions is not natural. Deep down they look three dimensional.

\newgeometry{left=0cm} 
\pagecolor{titlepagecolor}
\restoregeometry 
\end{document}